\documentclass[twocolumn,prb,aps,showpacs]{revtex4}

\usepackage{graphicx}
\usepackage{dcolumn}
\usepackage{amsmath}
\newlength{\figwidth}
\newlength{\figwidthb}
\begin{document}


\title{Resonant inelastic x-ray scattering study of the electronic structure of Cu$_2$O}
\author{Young-June Kim}
\email{yjkim@physics.utoronto.ca} \affiliation{Department of
Physics, University of Toronto, Toronto, Ontario M5S~1A7, Canada}
\author{J. P. Hill}
\affiliation{Department of Condensed Matter Physics and Materials
Science, Brookhaven National Laboratory, Upton, New York 11973}
\author{H. Yamaguchi}
\affiliation{Department of Electronics and Information Systems,
Akita Prefectural University, 84-4 Ebinokuchi Tsuchiya, Honjo, Akita
015-0055, Japan}
\author{T. Gog}
\affiliation{XOR, Advanced Photon Source, Argonne National
Laboratory, Argonne, Illinois 60439}
\author{D. Casa}
\affiliation{XOR, Advanced Photon Source, Argonne National
Laboratory, Argonne, Illinois 60439}

\date{\today}

\begin{abstract}
A resonant inelastic x-ray scattering study of the electronic
structure of the semiconductor cuprous oxide, $\rm Cu_2O$, is
reported. When the incident x-ray energy is tuned to the Cu
K-absorption edge, large enhancements of the spectral features
corresponding to the electronic transitions between the valence band
and the conduction band are observed. A feature at 6.5 eV can be
well described by an interband transition from occupied states of
mostly Cu 3d charactor to unoccupied states with mixed 3d, 4s and 2p
character. In addition, an insulating band gap is observed, and the
momentum dependence of the lower bound is measured along the
$\Gamma-R$ direction. This is found to be in good agreement with the
valence band dispersion measured with angle-resolved photoemission
spectroscopy.

\end{abstract}


\maketitle

\section{introduction}

In recent years, resonant inelastic x-ray scattering (RIXS) has been
used as a probe of various types of electronic excitations in many
condensed matter systems.\cite{Kotani01} In these studies, when the
incident photon energy is tuned to the x-ray absorption edge of
interest, the intensity of certain spectral features is greatly
enhanced, sometimes by up to a few orders of magnitude. Initially,
soft x-ray photons were used to study the electronic structure in
semiconductors and insulators.\cite{Kotani01} Recently, RIXS at
transition metal K edges has also been gaining interest due to the
added benefit of the momentum-resolving capability provided by the
short wavelength of the hard x-ray
photons.\cite{Hasan00,LCO-PRL,112-RIXS,Ishii05b} Such transition
metal K-edge RIXS is often called {\em indirect} RIXS, since the
intermediate state does not directly involve 3d photoelectrons,
unlike RIXS in the soft x-ray regime (transition metal L-edges),
which is therefore referred to as {\em direct}.

One of the attractions presented by hard x-ray RIXS is the
possibility of its use in studying momentum-resolved electronic
structure. In a simple picture of the RIXS cross-section, the
momentum-dependent part of the cross-section may be expressed as the
joint density of states (JDOS) of the unoccupied and occupied bands
for a particular momentum
transfer.\cite{Kotani01,Schulke-book,xsection} In other words, the
RIXS spectrum can be regarded as measuring inter-band transitions at
a finite momentum transfer, with an appropriate prefactor. Although
this viewpoint is probably too simplistic, attempts to develop a
linear response theory for the RIXS cross-section based on this
approach have been
made.\cite{Nomura05,vandenBrink06,Ament07,Takahashi08} For example,
in the limit of either very weak or very strong core hole potential,
van den Brink and van Veenendaal have shown that the indirect RIXS
cross section is a linear combination of the charge response
function and the dynamic longitudinal spin density correlation
function.\cite{vandenBrink06} This line of approach is quite
intriguing, since it suggests that one could use RIXS as a tool for
reconstructing electronic structure at the finite momentum
transfer.\cite{footnote1}

Despite extensive RIXS investigations of strongly correlated cuprate
compounds, little work has been done to date to study conventional
semiconductors with well-known electronic structure. This is
unfortunate, because investigation of well-characterized materials
with a known band structure is essential to formulate quantitative
understanding of this new experimental technique. In this paper, we
present just such an investigation of semiconducting $\rm Cu_2O$,
using the Cu K-edge RIXS technique, and show that the indirect RIXS
can measure JDOS of this weakly correlated system.

Cuprous oxide, $\rm Cu_2O$, is a naturally occurring mineral called
``cuprite." It is a direct-gap semiconductor with well-characterized
exciton lines, which are often considered as textbook examples of
Mott-Wannier type excitons.\cite{Kittel95} In particular,
Bose-Einstein condensation of the exciton gas has been reported and
has drawn much attention over the last two
decades.\cite{Snoke87,Wolfe95} In recent years, however, this
material has been drawing renewed interest due to its potential
applications in solar energy conversion and catalysis. \cite{Gou03}
Considerable attention has also been paid to this material due to
the unusual O-Cu-O linear bonding. \cite{Zuo99}

Cuprous oxide crystallizes in a cubic structure with space group
$Pn\bar{3}m$.\cite{Restori86,Eichhorn88} Since Hayashi and Katsuki's
measurement of exciton absorption lines,\cite{Hayashi50} early
experimental work on this material focused mostly on optical
absorption studies of the different exciton series, and is reviewed
in Refs.~\onlinecite{Nikitine69,Gross56}. Further, the electronic
structure of $\rm Cu_2O$ has been studied using various {\it ab
initio}
methods.\cite{Dahl66,Kleinman80,Ching89,Ruiz97,Nie02,Laskowski03,Filippetti05,Bruneval06,Hu08}
These studies suggest that $\rm Cu_2O$ is a semiconductor with
direct band gap, although the calculated gap varies widely among the
different approaches, and typically disagrees with the experimental
value of about 2 eV. (For a comprehensive comparison, see Table I in
Ref.~\onlinecite{Hu08}).

Advances in core level spectroscopy techniques developed at
synchrotron x-ray sources have made it possible to test some of the
earlier predictions concerning the electronic structure of $\rm
Cu_2O$. X-ray absorption studies near copper L-edges\cite{Hulbert84}
as well as at the O K-edge \cite{Grioni92} have been carried out to
elucidate the unoccupied bands. Ghijsen and coworkers carried out
both x-ray photoemission spectroscopy (XPS) and bremsstrahlung
isochromat spectroscopy (BIS) to probe both valence band and
conduction band density of states, respectively.\cite{Ghijsen88}
These experimental results could be largely accounted for with the
calculated band structure, confirming that $\rm Cu_2O$ is a
conventional band insulator with no, or very weak, electron
correlation. The top of the valence band is dominated by Cu $d$
states, while the lowest lying conduction band is primarily of
$d$-character which hybridizes with Cu $4s$ and $4p$ states, and O
$2p$ states.

As a result of the renewed interest in this material, there have
been several recent studies of the electronic structure of $\rm
Cu_2O$ using state-of-the-art electron spectroscopy techniques.
Bruneval and coworkers measured the valence band dispersion along
the $\Gamma-R$ direction with the angle resolved photoemisson
spectroscopy (ARPES), and were able to explain the observed
dispersion using a self-consistent GW method,\cite{Bruneval06} while
\"Onsten et al. used high-energy ARPES to measure the valence band
dispersion along the $\Gamma-M$ direction.\cite{Onsten07} Finally,
and of particular interest here, Hu et al. carried out comprehensive
soft x-ray (direct) RIXS studies of $\rm Cu_2O$, and have compared
their results with a calculation based on JDOS-like interband
transitions.\cite{Hu08}

In this paper, we present a detailed study of $\rm Cu_2O$ using
indirect RIXS. In particular, we will focus on the energy region
near the band gap, and report our measurement of the dispersion
relation of the band-gap edge along the $\Gamma-R$ direction. The
observed momentum dependence of the gap edge is consistent with the
dispersion of the valence band as observed with
ARPES.\cite{Bruneval06} In addition, we observe another band of
excitations at 6.5 eV, which can be explained as interband
transitions from the occupied Cu 3d state to the unoccupied state,
which is a mixture of Cu 3d, 4s, and O 2p. We contrast this with the
soft RIXS data of Hu et al.\cite{Hu08} Taken together, these results
suggest that the indirect (Cu K-edge) RIXS is a good probe of
electronic structure.

\section{experimental details}
\label{sec:exp}

The RIXS experiments were carried out at the Advanced Photon Source
on the undulator beamline 9IDB. A double-bounce Si(333)
monochromator and a spherical (R=1~m), diced  Ge(733) analyzer was
used to obtain an overall energy resolution of 0.4 eV (FWHM). We
also utilized a high-resolution setup with a Si(444) channel-cut
secondary monochromator and a diced Ge(733) analyzer with 2~m radius
of curvature. This provides an overall energy resolution of 0.13~eV.

A single crystal sample of $\rm Cu_2O$ was grown using the traveling
solvent floating zone method. The details of the crystal growth was
reported in Ref.~\onlinecite{Ito98b}. The crystal used in the RIXS
measurements came from the same batch as the one used in the optical
studies reported in Ref.~\onlinecite{Ito98a}. The as-grown crystal
was mechanically polished and etched in order to optimize the
surface condition. The crystal was mounted on an aluminium sample
holder at room temperature inside an evacuated chamber.

Most of the measurements were carried out near the {\bf Q}=(2,0,1)
point, which also corresponds to a reciprocal wave vector {\bf G}.
The reduced momentum transfer {\bf q} is defined as ${\bf q} \equiv
{\bf Q}-{\bf G}$, and is measured from the Brillouin zone center.
Note that the $\rm Cu_2O$ structure is composed of an
interpenetrating fcc lattice of Cu and a bcc lattice of O.
Therefore, $(h, k, l)$ reflections with unmixed indices ($h,k,l$ all
even or all odd) will have a strong Bragg peak intensity, following
the fcc structure factor, while the reflections with mixed indices,
but $h+k+l$=even, will have a weak Bragg peak intensity coming from
the bcc lattice of oxygens. Reflections with $h+k+l$=odd, such as
the (2,0,1) point, are forbidden. However, Eichhorn and coworkers
reported intensity at such positions with the incident energy near
the Cu K-edge, due to the anisotropic Cu
environment.\cite{Eichhorn88} As a result, the ${\bf q}=0$ point
data were in fact taken at the (2,0.05,1) position, to reduce the
background due to the resonant elastic intensity.

\section{results and discussion}

\subsection{Energy dependence}
\label{sec:res1}

\begin{figure}
\begin{center}
\includegraphics[width=\figwidth]{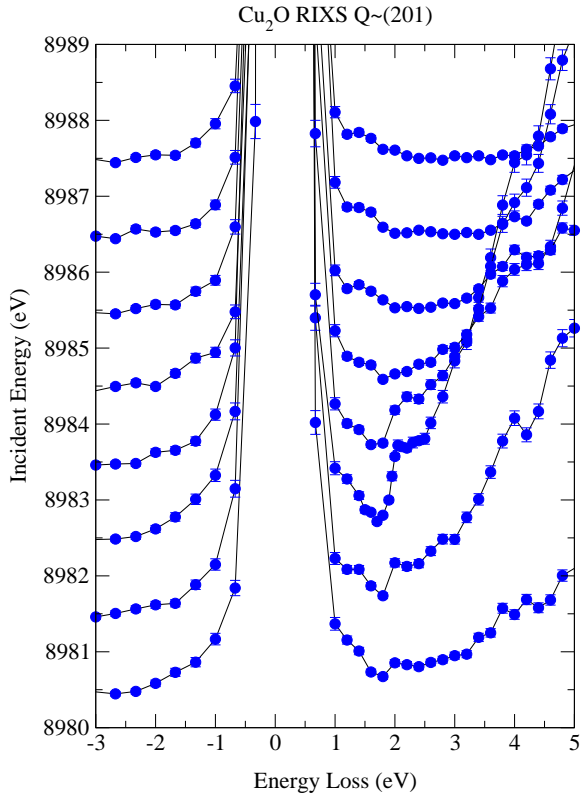}
\end{center}
\caption{Scattered intensity at {\bf Q}$\approx$(2 0 1), plotted as
a function of energy loss, $\hbar \omega$. The scans are shifted
vertically for clarity. The incident energy for each scan can be
read off from the vertical axis.} \label{fig:resonance}
\end{figure}

In Fig.~\ref{fig:resonance}, we show representative RIXS scans
obtained with different incident photon energies ($E_i$) as denoted
on the left side of the figure. Each scan is taken at the (2,0.05,1)
position, and plotted as a function of energy loss ($\hbar
\omega=E_i - E_f$). In each scan, there is a large zero energy loss
peak arising from the quasi-elastic background due to phonons and
static disorder. In addition, there is a broad, strong feature on
the high energy side of the scan. Between these two prominent
features, one can observe a sharp edge-like RIXS feature around
$\hbar \omega \approx 2$ eV, which exhibits a large resonant
enhancement at $E_i=8982.5$ eV.

In general, there are two distinct types of spectral features
observed by RIXS: One is a valence electron excitation that appears
at fixed energy loss ($\hbar \omega$) regardless of the incident
energy, similar to Raman scattering peaks. The sharp edge-like
feature at 2 eV corresponds to such an excitation. The other type is
due to emission lines arising from radiative transitions between
atomic-like states, which are observed at fixed final photon energy
($E_f$). When plotted as a function of energy loss, as shown in
Fig.~\ref{fig:resonance}, the peak position of the latter features
therefore shifts as the incident energy is varied above the edge.
The broad feature at higher energy ($\sim 5$ eV) in
Fig.~\ref{fig:resonance} corresponds to such a feature.

The sharp edge-like feature shows quite a narrow resonance behavior
around the incident energy of $E_i=8982.5$ eV. We note that this
energy corresponds to the sharp peak in the x-ray absorption spectra
(XAS) shown in Fig.~8 of Ref.~\onlinecite{Lytle88} and Fig.~8 of
Ref.~\onlinecite{Guo90}. Since this is the strongest absorption
feature, we can associate this incident energy, and the
corresponding intermediate state of the resonant inelastic process,
with the $\underline{1s}4p$ state following a $1s \rightarrow 4p$
dipole transition, where $\underline{1s}$ denotes a $1s$ core hole.
Since Cu is found with Cu$^+$ valence in $\rm Cu_2O$, this
intermediate state energy is shifted down by about 10 eV compared to
the Cu$^{2+}$ state in cuprate materials like $\rm
La_2CuO_4$.\cite{LCO-PRL} Note that the RIXS spectra showed no
dependence on the incident photon polarization, presumably due to
the cubic symmetry of the sample. We identify this sharp edge-like
feature as the {\bf q}=0 band gap, since its energy-loss of $\hbar
\omega \approx 2$ eV coincides with the optical gap energy. As we
will see in the next section, this identification is corroborated by
the momentum dependence of this feature.

\begin{figure}
\begin{center}
\includegraphics[width=3.3in]{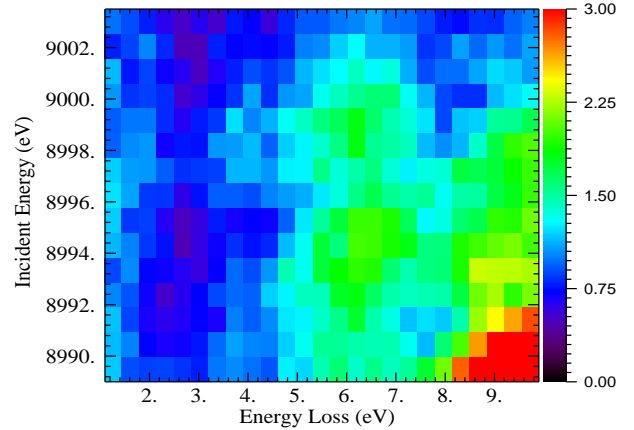}
\end{center}
\caption{(Color online) Scattered intensity at {\bf Q}$\approx$(2 0
1), plotted in pseudo-color intensity map. The horizontal axis is
Energy loss and the vertical axis the incident photon energy. The
intensity scale bar is shown on the right. Note that the incident
energies are higher than those in Fig. 1.} \label{fig:res2}
\end{figure}

\begin{figure}
\begin{center}
\includegraphics[width=\figwidth]{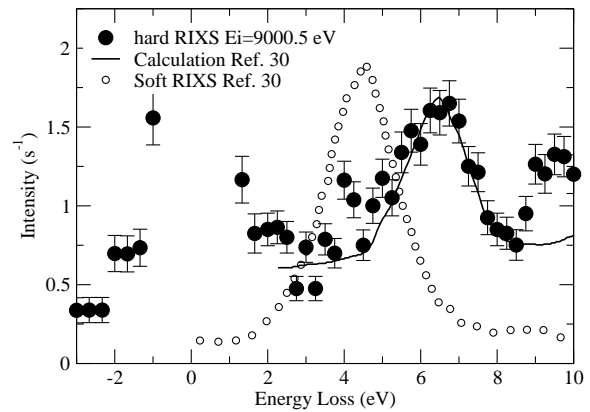}
\end{center}
\caption{Energy loss scan obtained with $E_i=9000.5$ eV. The
momentum is {\bf Q}$\sim$(2 0 1). The open circles are experimental
soft x-ray RIXS spectrum from Ref.~\onlinecite{Hu08}. The solid line
is a calculated RIXS spectrum from Ref.~\onlinecite{Hu08} scaled to
match the current data.} \label{fig:soft}
\end{figure}

In addition to this low energy feature at the band gap, there are
higher energy features, which show distinct incident energy
dependence. In Fig.~\ref{fig:res2}, we show an intensity map as a
function of both incident energy and energy loss again taken at
(2,0.05,1). Note that the incident energies used in
Fig.~\ref{fig:res2} are higher than those for the data shown in
Fig.~\ref{fig:resonance}. One can clearly identify a very broad and
strong spectral feature located at around $\hbar \omega \approx 6.5$
eV, which is resonantly enhanced over a wide range of incident
energies; most notably around $E_i=8994$ eV and $E_i=8999$ eV. In
addition, there is a higher energy feature at $\hbar \omega \approx
9-10$ eV, and a weak low energy feature located around $\hbar \omega
\approx 4$ eV that resonates around $E_i=8999$ eV. Since the peak
positions of the loss features do not change with the incident
energy, we chose to examine the $E_i=9000.5$ eV scan, which exhibit
all three features in more detail. These data are shown in
Fig.~\ref{fig:soft}.

\begin{figure}
\begin{center}
\includegraphics[width=3.3in]{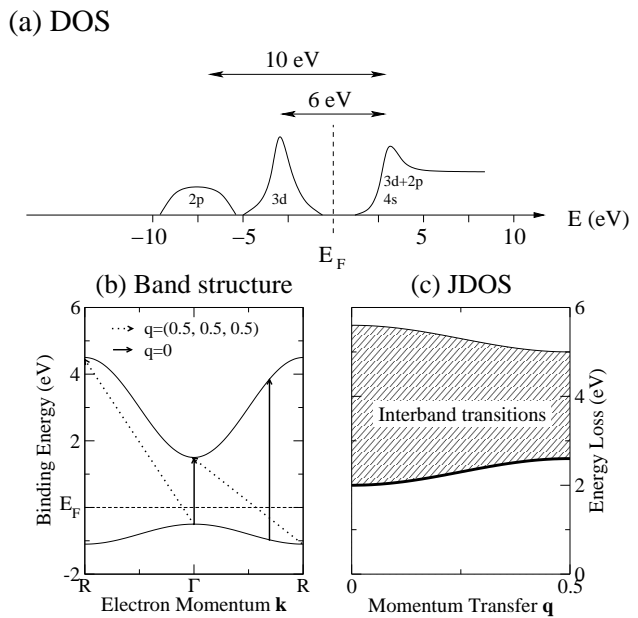}
\includegraphics[width=\figwidth]{fig4b.eps}
\end{center}
\caption{(a) Schematic representation of experimental density of
states (DOS) from Ref.~\onlinecite{Ghijsen88}. (b) Simplified band
structure shown for illustrative purposes as described in the text.
The arrows denote possible interband transitions with different
momentum transfers. Vertical solid lines are ${\bf q}=0$
transitions; dashed lines correspond to ${\bf q}=(1/2, 1/2, 1/2)$
transitions. (c) Momentum dependence of the joint density of states
(JDOS) arising from the band structure shown in (b). Note that the
magnitude of the momentum dependence of the lower bound of the
interband transition continuum (thick solid line) mimics that of the
valence band.} \label{fig:dos}
\end{figure}

To understand the origin of these excitations, it is useful to
consider the electronic density of states. Using both XPS and BIS
techniques, Ghijsen {\it et al.} \cite{Ghijsen88} reported valence
band and conduction band density of states for $\rm Cu_2O$. The most
prominent peak in BIS occurs around 3 eV above the Fermi level,
while XPS shows a large peak centered around 3 eV below the Fermi
level with shoulders on both sides. The density of states based on
these XPS and BIS results are shown schematically in
Fig.~\ref{fig:dos}(a). The main peak in the XPS spectra is mainly of
Cu d-character (based on the various calculations), while the larger
binding energy shoulder, at about 7 eV below the Fermi level, has
mainly O $2p$ character.\cite{Ghijsen88,Ching89,Ruiz97} The nature
of the low-lying unoccupied states has been attributed to mixed
$3d-2p$ states\cite{Ghijsen88} or to Cu $4s$ states.\cite{Ruiz97}.
Given this picture, it is natural to associate the $\hbar \omega
=6.5$ eV RIXS feature with the transition from the $3d$ to either
$3d-2p$ mixture or $4s$. Then the higher energy RIXS feature at
9-10~eV presumably is due to the transition from the occupied O $2p$
to empty $3d$ or $4s$ states.

To make a more quantitative comparison, the hard RIXS scan obtained
with $E_i=9000.5$ eV is compared with the calculated spectrum, taken
from Ref.~\onlinecite{Hu08}, in Fig.~\ref{fig:soft}. Hu and
coworkers computed the inelastic x-ray scattering intensity due to
dipole allowed interband transitions between filled and empty
partial density of states.\cite{Hu08} The most dominant loss
structure is found around 6.5 eV through the $3d \rightarrow 4s/3d$
channels, while there exist weaker features at 4 eV and 10 eV in the
other channels ($4s \rightarrow 4s/3d$). Since the matrix elements
for the K-edge experiments are unknown, we plot simply the dominant
$3d \rightarrow 4s$ transition (solid line) without any matrix
element weighting in Fig.~\ref{fig:soft}. We find it describes the
main peak at 6.5 eV remarkably well. Note that neither the peak
position nor the core-hole broadening (1.35 eV) has been altered
from the calculation in Ref.~\onlinecite{Hu08}, and only the
intensity scale has been changed. On the other hand, the Cu $L_3$
RIXS spectrum obtained with $E_i=933.75$ eV in
Ref.~\onlinecite{Hu08} exhibits a very different shape, with an
energy loss feature centered around 4.5 eV, as shown in
Fig.~\ref{fig:soft} (open circles). Hu and coworkers used a
phenomenological excitonic density of states in order to account for
this discrepancy. This was an ad-hoc approach that used the
experimental L-edge absorption spectrum (for which there is a 2p
core hole in the final state) in place of the calculated occupied
DOS. This was labeled the ``excitonic DOS". Although there is no 2p
core hole in the L-edge RIXS final state, such excitonic DOS was
required to explain experimental observations. Figure~\ref{fig:soft}
shows that no such ad-hoc corrections are required to explain the
K-edge RIXS data, which are well reproduced by the theoretical JDOS.
Thus, the present comparison suggests that Cu K-edge RIXS (often
called indirect RIXS) can measure band structure directly, unlike Cu
$L_3$-edge RIXS. Note that the additional features observed at 4 eV
and 10 eV in the K-edge RIXS could also be explained by the $4s
\rightarrow 3d$ transition according to the calculation \cite{Hu08}.

\subsection{Momentum dependence}
\label{sec:res2}

\begin{figure}
\begin{center}
\includegraphics[width=\figwidth]{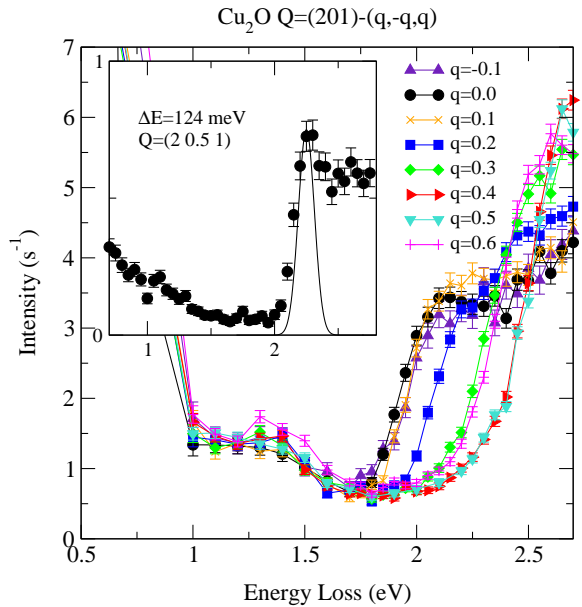}
\end{center}
\caption{(Color online) RIXS intensity at different Q positions
obtained with fixed incident energy at $E_i=8982.5$ eV. The Q
positions are along the $\Gamma-R$ line in reciprocal space, that
is, along the (1,1,1) direction. The inset shows the high-resolution
data. The resolution function of 124 meV full width is shown as a
solid line. Note that the weak feature at 1.3 eV observed in the
main figure is an experimental artifact, and is not observed in the
high-resolution data.} \label{fig:dispersion}
\end{figure}

We now turn to the 2 eV feature. With the incident energy fixed at
$E_i=8982.5$ eV, we studied the momentum dependence of the sharp
edge feature observed at 2~eV (Fig.~\ref{fig:resonance}). In
Fig.~\ref{fig:dispersion}, we plot the RIXS spectra obtained at {\bf
Q} positions along the $\Gamma-R$ line, that is, with {\bf q} along
the $(1,\bar{1},1)$ direction. We plot only a narrow frequency
range, in order to emphasize the evolution of the lowest energy
excitation as a function of momentum transfer. (Note that there
appears to be a small peak at $\hbar \omega \sim 1 -1.5 $eV.
However, this feature is present in all scans and has no $E_i$
dependence, nor is it observed in the high resolution data, as shown
in the inset of Fig.~\ref{fig:dispersion}. Thus, it is most likely
due to an experimental artifact, and we do not discuss it further.)

\begin{figure}
\begin{center}
\includegraphics[width=\figwidth]{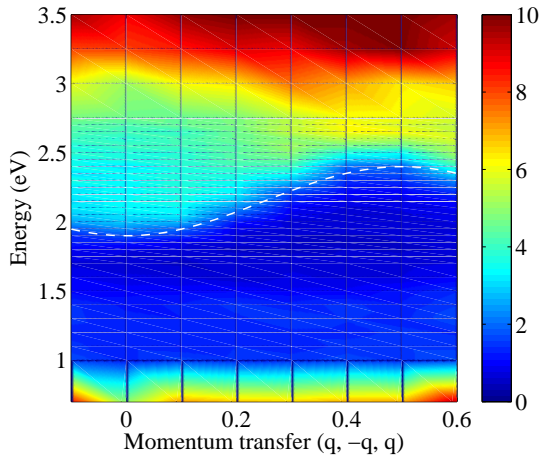}
\end{center}
\caption{(Color Online) Intensity colormap as a function of energy
and momentum transfer. The intensity scale in counts per second is
shown on the right hand side of the main figure. Intensity of 10
counts per second or higher is shown in brown.}
\label{fig:disp-color}
\end{figure}

The lowest energy excitation across the band gap shows quite a
dramatic momentum dependence. In order to show the dispersion of
this excitation graphically, we plot the intensity as a function of
$\hbar \omega$ and $q$ in Fig.~\ref{fig:disp-color}. The intensity
scale is shown on the right hand side, and is described in the
figure caption. The boundary between the high intensity region and
the background forms a sinusoidal dispersion relation of the form
$\hbar \omega \approx 2.15-0.25\cos(2\pi q)$, as shown as the white
dashed line in Fig.~\ref{fig:disp-color}.

If one adopts the simple view that the RIXS spectrum in a weakly
correlated system is proportional to the JDOS at a given momentum
transfer, then the observed dispersion in Fig.~\ref{fig:disp-color}
should reflect the momentum dependence of the lower bound for
inter-band transitions. In order to understand this, let's consider
a simple sinusoidal conduction and valence band, with a direct gap
located at the $\Gamma$ position. An idealized representation of
such a band structure is shown in Fig.~\ref{fig:dos}(b) as a
function of electron momentum {\bf k} along the $\Gamma-R$
direction. Here the valence band dispersion $E_v(k)$ is taken from
the recent angle-resolved photoemission (ARPES)
data,\cite{Bruneval06} while the conduction band dispersion $E_c(k)$
is taken from the calculation in Ref.~\onlinecite{Ruiz97}.

Finite {\bf q} interband transitions can be represented as vectors
connecting two points on the $E_v$ and $E_c$ curves as shown in
Fig.~\ref{fig:dos}(b). The frequency and momentum of a particular
transition satisfies $\hbar \omega ({\bf q}) = E_c({\bf k}_p) -
E_v({\bf k}_h)$ and ${\bf q}= {\bf k}_p - {\bf k}_h$, where ${\bf
k}_p$ and ${\bf k}_h$ are particle and hole momentum, respectively.
Examples of the ${\bf q}=0$ transition are shown in solid lines,
while the dotted line represents examples of the ${\bf
q}=(\frac{1}{2}, \frac{1}{2}, \frac{1}{2})$ transition. Integrating
over all possible ${\bf k}_p$ and ${\bf k}_h$ combinations, one can
obtain the ${\bf q}$-dependence of inter-band transitions along the
(111) direction. This is shown for this simplified band structure in
Fig.~\ref{fig:dos}(c). Since the conduction bandwidth is much larger
than the valence bandwidth in this case, the {\em lower bound} of
inter-band transitions is dominated by the valence band. In other
words, the lower bound transitions involve exciting electrons from
all ${\bf k}_h$ positions of the valence band to the bottom of the
conduction band (${\bf k}_p=0$), since it takes more energy to
excite electrons into finite ${\bf k}_p$ states in conduction band.
Therefore, we expect the dispersion of the valence band to be
similar to the dispersion of the gap observed in our RIXS
measurements. This expectation is borne out by the recent ARPES
studies by Bruneval et al. \cite{Bruneval06} and \"Onsten et al.
\cite{Onsten07}. In particular, the valence band dispersion along
the $\Gamma-R$ direction was reported in
Ref.~\onlinecite{Bruneval06}, which has the valence band maximum at
the $\Gamma$ position and bandwidth of about 0.6 eV as shown in
Fig.~\ref{fig:dos}(b). This is very similar to the bandwidth 0.5 eV
observed in Fig.~\ref{fig:disp-color}. Though, we should note that
the picture considered in Fig.~\ref{fig:dos}(b) is an
oversimplification in one dimension, and in order to make a
quantitative comparison, one would have to take into account the
full three dimensional band structure.

\section{conclusions}

A resonant inelastic x-ray scattering (RIXS) study of the electronic
structure of cuprous oxide, $\rm Cu_2O$, is reported. By tuning the
incident x-ray photon energy to the Cu K-absorption edge, resonantly
enhanced electronic transitions are observed. Comparison with Cu
$L_3$-edge RIXS spectrum and a calculation based on band-structure
density-of-states clearly illustrates that the indirect RIXS
spectrum of $\rm Cu_2O$ obtained at the Cu K-edge can be explained
simply by interband transitions. We also observe that the momentum
dependence of the lowest energy inter-band transitions along the
$\Gamma-R$ direction, and show that it can be well described by the
dispersion of the valence band of Cu $3d$ character.

Our main result is that the indirect RIXS cross-section seems to be
described by interband transitions from the occupied valence band to
the unoccupied conduction band. This gives credence to theoretical
efforts to connect the RIXS cross-section with the charge
correlation function,\cite{vandenBrink06} and also allows one to use
indirect RIXS to investigate the electronic structure of materials.
This last point is an important one, since RIXS has many advantages
over other electron spectroscopy techniques, such as
bulk-sensitivity, element-sensitivity, and momentum-resolving
capability.

\acknowledgements{We would like to thank J. van den Brink, W. Ku,
and M. van Veenendaal for invaluable discussions. The work at
University of Toronto was supported by Natural Sciences and
Engineering Research Council of Canada. The work at Brookhaven was
supported by the U. S. DOE, Office of Science Contract No.
DE-AC02-98CH10886. Use of the Advanced Photon Source was supported
by the U. S. DOE, Office of Science, Office of Basic Energy
Sciences, under Contract No. W-31-109-ENG-38.}

\end{document}